\documentclass[12pt]{article} 
\usepackage{graphicx}
\usepackage{amsbsy}

\begin{document}
\baselineskip 10mm

\vskip 2mm

\centerline{\large \bf On the analogy between the classical wave optics}
\centerline{\large \bf and the quantum wave phenomena}

\vskip 2mm

\centerline{L. A. Openov}

\vskip 2mm

\centerline{\it Moscow Engineering Physics Institute
(State University)}
\centerline{\it 115409 Moscow, Russia}
\centerline{\it E-mail: opn@supercon.mephi.ru}

\vskip 4mm

\begin{quotation}

A striking correspondence between the effects of an auxiliary-mode-assisted
transfer of light power between two waveguides and an
auxiliary-state-assisted transfer of an electron between two quantum dots is
highlighted by the example of an exactly solvable model.

\end{quotation}

\vskip 4mm


\vskip 6mm

\vskip 6mm

It has long been known that sometimes the fundamentally different physical
phenomena can be described with a common mathematical apparatus. An
illustrative example is application of methods of the quantum field theory
in condensed matter physics \cite{AGD}. Another example is some kind of
analogy between the classical electromagnetic wave optics and the quantum
wave phenomena \cite{Kawano}. Recently, Vorobeichik {\it et al.}
reported the results of experimental study on the light power transfer
between two parallel optical waveguides \cite{Vorobeichik}.
They have shown that in the presence of periodic
longitudinal spatial perturbation of the waveguides, the power transfer is
enhanced by about three orders of magnitude. According to
their earlier work \cite{Vorobeichik2}, the power
exchange is due to interaction of lowest order waveguide modes with
auxiliary high order modes. To analyze the experimental findings, the authors
of Refs. \cite{Vorobeichik,Vorobeichik2} made use of the formal analogy
between the Maxwell and Schr\"odinger equations \cite{Kawano}.

In this Letter,
by the example of an exactly solvable model, we show that there is not only
the mere analogy but a striking one-to-one correspondence between the effect
of auxiliary-mode-assisted transfer of light power from the lowest order mode
of one waveguide to the lowest order mode of another waveguide and the effect
of auxiliary-state-assisted transfer of an electron from the lowest localized
state of one quantum dot to the lowest localized state of another quantum
dot, the latter being equivalent to the effect of auxiliary-level-assisted
coupling of two logical states of the superconducting Josephson-phase qubit
suggested to realize a general quantum gate without tunneling \cite{Amin}.

In weakly guiding structures with a slow $z$-dependence of the refractive
index $n(\pmb{\rho},z)$, the Maxwell equation
\begin{equation}
\Delta E(\pmb{\rho},z)+k^2(\pmb{\rho},z) E(\pmb{\rho},z)=0
\label{Maxwell}
\end{equation}
for the nonzero component $E(\pmb{\rho},z)$ of a linearly polarized electric
field can be rewritten \cite{Vorobeichik,Vorobeichik2} by the substitution
\begin{equation}
E(\pmb{\rho},z)=\Phi(\pmb{\rho},z)\exp(ik_0z)
\label{Substitution}
\end{equation}
as
\begin{equation}
\left[-\frac{\hbar_e^2}{2}\nabla_{\pmb{\rho}}^2+D(\pmb{\rho},z)\right]
\Phi(\pmb{\rho},z)=i\hbar_e\frac{\partial\Phi(\pmb{\rho},z)}{\partial z}~,
\label{Paraxial}
\end{equation}
where $\pmb{\rho}=(x,y)$,
$k(\pmb{\rho},z)=2\pi n(\pmb{\rho},z)/\lambda$, $k_0=2\pi n_0/\lambda$, $n_0$
is the refractive index of the waveguide cladding, $\lambda$ is the free
space wavelength, $\hbar_e=1/k_0$, and
\begin{equation}
D(\pmb{\rho},z)=\frac{1}{2}
\left[1-\left(\frac{n(\pmb{\rho},z)}{n_0}\right)^2\right]~.
\label{D(rho,z}
\end{equation}
The equation (\ref{Paraxial}) for
$\Phi(\pmb{\rho},z)$ is formally equivalent to the Schr\"odinger equation
\begin{equation}
\left[-\frac{\hbar^2}{2m^*}\nabla_{\bf r}^2+U({\bf r},t)\right]
\Psi({\bf r},t)=i\hbar\frac{\partial\Psi({\bf r},t)}{\partial t}
\label{Schrodinger}
\end{equation}
for the wave function
$\Psi({\bf r},t)$ of an electron moving in the time-dependent potential
$U({\bf r},t)$, where ${\bf r}=(x,y,z)$, $m^*$ is the effective electron
mass, and $\hbar$ is the Planck constant. This equation turns into equation
(\ref{Paraxial})
for the classical electric field at ${\bf r}\rightarrow\pmb{\rho}$,
$t\rightarrow z$, $m^*\rightarrow 1$, $\hbar\rightarrow\hbar_e$,
$V({\bf r},t)\rightarrow D(\pmb{\rho},z)$, and
$\Psi({\bf r},t)\rightarrow\Phi(\pmb{\rho},z)$. The eigenenergies
$\varepsilon_n$ of the stationary Schr\"odinger equation for the
eigenfunctions
$\Psi_n({\bf r})$ correspond to the eigenvalues $d_n$ of the equation for
ideal modes $\Phi_n(\pmb{\rho})$ in the unperturbed waveguide.

The operation of directional couplers is based on beatings between the modal
fields of two parallel dielectric waveguides \cite{Snyder}, so that the
input light power localized in one of the waveguides is transferred
completely to another waveguide. Such beatings are the optical analog of the
quantum tunneling phenomena. The complete exchange of the light power between
the two waveguides takes place at the beating length $l_b$. Typically,
$l_b\sim 1$ cm. Increase in the waveguide separation leads to exponential
increase in $l_b$, so that the directional coupling becomes unobservable.
The authors of Refs. \cite{Vorobeichik,Vorobeichik2} succeeded in the
orders-of-magnitude reduction of $l_b$ in such structures through spatial
variations of the waveguide geometry. Here we show that this effect is
analogous to the fast electron transfer between the quantum dots under the
influence of the resonant electromagnetic field \cite{Openov}.

Let us consider a nanostructure composed of two identical semiconductor
quantum dots L and R, see Fig. 1. An extra electron occupies initially
the lowest size-quantized state $|L\rangle$ with the energy $\varepsilon_0$
in the conduction band of the dot L. The lowest state $|R\rangle$ of the dot
R has the same energy and is unoccupied at $t=0$. The wave functions
$\langle{\bf r}|L\rangle$ and $\langle{\bf r}|R\rangle$ are localized in the
dots L and R, respectively. If the energy $\Delta$ of electron tunneling
between the two dots is much smaller than the energy separation between the
states $\langle{\bf r}|L,R\rangle$ and the nearest excited state of the
nanostructure, the electron Hamiltonian can be written as
\begin{equation}
\hat{H}_0=\varepsilon_0\bigl(|L\rangle\langle L|+|R\rangle\langle R|\bigr)-
\Delta\bigl(|L\rangle\langle R|+|R\rangle\langle L|\bigr)~.
\label{H_0}
\end{equation}
The states $|L\rangle$ and $|R\rangle$ can be represented as symmetric and
antisymmetric superpositions of the eigenstates $|1\rangle$ and $|2\rangle$
with the eigenenergies $\varepsilon_1=\varepsilon_0-\Delta$ and
$\varepsilon_2=\varepsilon_0+\Delta$, respectively, as
$|L\rangle=[|1\rangle+|2\rangle]/\sqrt{2}$ and
$|R\rangle=[|1\rangle-|2\rangle]/\sqrt{2}$.
Since the states $|L\rangle$ and $|R\rangle$ are not the exact eigenstates of
the Hamiltonian $\hat{H}_0$, the initial state $|\Psi(0)\rangle=|L\rangle$
will evolve with time as
\begin{equation}
|\Psi (t)\rangle = e^{-i\hat{H}_0 t/\hbar} |\Psi (0)\rangle =
e^{-i(\varepsilon_0-\Delta) t/\hbar}\biggl\{ |L\rangle -
ie^{-i\Delta t/\hbar}
\sin(\Delta t/\hbar)\biggl[|L\rangle-|R\rangle\biggr]\biggr\} ~,
\label{Psi(t)}
\end{equation}
so that the electron will tunnel to the dot R in time $T=\hbar\pi/2\Delta$.
In the case that the energy barrier $U_b$ separating the dots is high and/or
the ratio of the dot spacing $d$ to the dot size $a$ is large, the value of
$\Delta$ appears to be exponentially small, so that
$\varepsilon_1\approx\varepsilon_2\approx\varepsilon_0$, and the electron remains
localized in the dot L for a macroscopically long time, e. g.,
$T\sim 10^8$ s at $U_b=1$ eV and $d/a=3$, see Ref. \cite{Openov}.

In order to facilitate the fast electron transfer between the dots of such
a nanostructure, one can make use of the external electromagnetic field
\cite{Openov}. We assume that there is an excited level $|3\rangle$ in the
nanostructure (not necessarily third in order), whose energy $\varepsilon_3$ is
close to the top of the potential barrier separating the dots, so that the
corresponding wave function $\langle{\bf r}|3\rangle$ is delocalized between
the dots. Let the nanostructure be subjected to a periodic stepwise
perturbation
\begin{equation}
V({\bf r},t)=V({\bf r})f(t)=
V({\bf r})\sum_{j=1}^N\theta(t+\frac{T_0}{2}-jT_0)\theta(jT_0-t)
\label{V(r,t}
\end{equation}
with the period $T_0$ and the duration $T=NT_0>>T_0$. If the value of
$(\varepsilon_3-\varepsilon_1)/\hbar\approx(\varepsilon_3-\varepsilon_2)/\hbar$
equals to
one of the frequencies $\Omega_n=2\pi n/T_0$ in the Fourier expansion
\begin{equation}
f(t)=\sum_k f(\Omega_k)\exp(-i\Omega_k t)~,
\label{f(t)}
\end{equation}
where
\begin{equation}
f(\Omega_k)=\frac{1}{T_0}\int_0^{T_0}dtf(t)\exp(i\Omega_k t)=
\frac{\sin\left(\frac{\pi k}{2}\right)}{\pi k}
\exp\left(-\frac{i\pi k}{2}\right)~,
\label{f(omega)}
\end{equation}
then both states $|1\rangle$ and $|2\rangle$ are resonantly coupled to the
state $|3\rangle$. This coupling results in the coherent evolution of the
electron state vector, so that the probability $p_R$ to find an electron in
the state $|R\rangle$ varies with time as \cite{Openov}
\begin{equation}
p_R(t)=|\langle R|\Psi(t)\rangle|^2=\sin^4(\omega t)~,
\label{pR(t)}
\end{equation}
where
\begin{equation}
\omega=|\langle R|V({\bf r})|3\rangle|\frac{|f(\Omega_n)|}{\hbar\sqrt{2}}~.
\label{omega}
\end{equation}
If $T=\pi/2\omega+\pi m/\omega$, where $m\geq 0$ is an integer, then
$p_R(T)=1$ and, hence, after the applied perturbation is off at $t=T$, the
electron will stay localized in the state $|R\rangle$. The value of $T$ can
be made many orders of magnitude shorter than in the case of direct electron
tunneling between the dots. For example, if the external perturbation is
associated with the electric field, $V({\bf r})=-e{\bf E_0 r}$, then
$T\sim \hbar/eE_0a \sim 10^{-8}$ s at $E_0=1$ V/cm and $a=1$ nm.

Now we turn to the double-waveguide structure. In Ref. \cite{Vorobeichik},
the periodic perturbation in the $z$-direction was generated by changing the
width of the parallel waveguides by $b=0.5$ $\mu$m in a stepwise fashion,
see Fig. 2, so that
\begin{equation}
\Delta n(\pmb{\rho},z)=(n_0-n_1)g(\pmb{\rho})f(z)~,
\label{Delta_n}
\end{equation}
where
$n_1$ is the refractive index of the waveguide cores, $g(\pmb{\rho})=1$ if
$\pmb{\rho}$ is within the distance $b$ of the corresponding waveguide side
and $g(\pmb{\rho})=0$ otherwise, and
\begin{equation}
f(z)=\sum_{j=1}^N\theta(z+\frac{L_0}{2}-jL_0)\theta(jL_0-z)~.
\label{f(z)}
\end{equation}
Here the period $L_0$ and the total length $L=NL_0$ of the spatial
perturbation correspond, respectively, to the period $T_0$ and the duration
$T$ of the time-dependent perturbation in the case of the resonant electron
transfer between the quantum dots. The periodic $z$-dependent change in
$D(\pmb{\rho})$ is
\begin{equation}
\Delta D(\pmb{\rho},z)=-\frac{\Delta n(\pmb{\rho},z)(n_0+n_1)}{2n_0^2}=
\Delta D(\pmb{\rho})f(z)~,
\label{D(z)}
\end{equation}
where
\begin{equation}
\Delta D(\pmb{\rho})=\frac{g(\pmb{\rho})}{2}
\left[\left(\frac{n_1}{n_0}\right)^2-1\right]~.
\label{DeltaD(z)}
\end{equation}
If the input light power is localized in the left waveguide, then
$\Phi(\pmb{\rho},z=0)=\Phi_L(\pmb{\rho})$, where
$\Phi_L(\pmb{\rho})=[\Phi_1(\pmb{\rho})+\Phi_2(\pmb{\rho})]/\sqrt{2}$ is the
symmetric superposition of the two lowest order trapped modes,
$\Phi_1(\pmb{\rho})$ and $\Phi_2(\pmb{\rho})$, of the unperturbed
double-waveguide structure, the corresponding eigenvalues of the
Schr\"odinger-like equation being equal to $d_1$ and $d_2$, respectively
($d_2\approx d_1$ for relatively large waveguide separation).
Let $\Phi_3(\pmb{\rho})$ be such a high-order mode with a
corresponding eigenvalue $d_3$ that is spread over the double-waveguide
structure and the value of $(d_3-d_1)/\hbar_e\approx (d_3-d_2)/\hbar_e$
equals to one of the wave
vectors $G_n=2\pi n/L_0$ in the Fourier expansion
\begin{equation}
f(z)=\sum_kf(G_k)\exp(-iG_kz)~,
\label{f(z)Fourier}
\end{equation}
where
\begin{equation}
f(G_k)=\frac{1}{L_0}\int_0^{L_0}dz f(z)\exp(iG_k z)=
\frac{\sin\left(\frac{\pi k}{2}\right)}{\pi k}
\exp\left(-\frac{i\pi k}{2}\right)~,
\label{f(G)}
\end{equation}

Making use of the
above mentioned correspondence between the equations for $\Phi(\pmb{\rho},z)$
and $\Psi({\bf r},t)$, one has for the directional coupling probability of
the light power
\begin{equation}
P_R(z)=|\langle R|\Phi(z)\rangle|^2=\sin^4(K z)~,
\label{pR(z)}
\end{equation}
where
\begin{equation}
K=|\langle R|\Delta D(\pmb{\rho})|3\rangle|\frac{|f(G_n)|}{\hbar_e\sqrt{2}}~.
\label{K}
\end{equation}
Here $|R\rangle$ and $|3\rangle$ denote the mode
$\Phi_R(\pmb{\rho})=[\Phi_1(\pmb{\rho})-\Phi_2(\pmb{\rho})]/\sqrt{2}$
localized in the right waveguide and the delocalized mode
$\Phi_3(\pmb{\rho})$, respectively. If $L=\pi/2K+\pi m/K$, where $m\geq 0$ is
an integer, then $P_R(L)=1$ and, hence, the output light power at $z=L$ will
be localized in the right waveguide. It is this effect that has been observed
in Ref. \cite{Vorobeichik}. For the waveguide structure studied in
Ref. \cite{Vorobeichik} a rough estimate gives $L=(1\div 10)$ mm, in a
qualitative agreement with the length of the periodic structure 7 mm in
Ref. \cite{Vorobeichik}. For the quantitative calculations of the spectrum
of the waveguide modes, one should
account for all specific details of the structure.

Finally, we note that the most probable reason for incomplete (about $50 \%$)
power transfer to the right waveguide and the excitement of different
high-order optical modes \cite{Vorobeichik} is the deviation from the
"resonant conditions" due to non-optimal waveguide structure parameters.
For example, in the case of the laser-induced electron transfer between the
quantum dots, when the frequency is offset from
resonance [$\delta=\Omega_n-(\varepsilon_3-\varepsilon_1)/\hbar\neq 0)$], the
maximum value of $p_R(t)$ is less than unity, e. g.,
$p_R(T)=1-(\pi^2/64)(\delta/\omega)^2$ at $T=\pi/2\omega$, see
Ref. \cite{Openov}. The difference in the dot sizes results in $p_R(T)$
decrease as well \cite{Tsukanov}. All expressions derived in
Refs. \cite{Openov,Tsukanov} for the ac field perturbed double-dot
nanostructure have their analogues in the spatially perturbed
double-waveguide dielectric structure and can be used for its engineering.

To conclude, the only essential difference between the two effects considered
is that one of them has been observed experimentally \cite{Vorobeichik},
while another \cite{Amin,Openov}, to the best of our knowledge, not yet.
We hope that this Letter will give impetus to experimental research in this
direction.

\vskip 2mm

I am grateful to Sergey Openov for assistance.

\newpage

\includegraphics[width=\hsize]{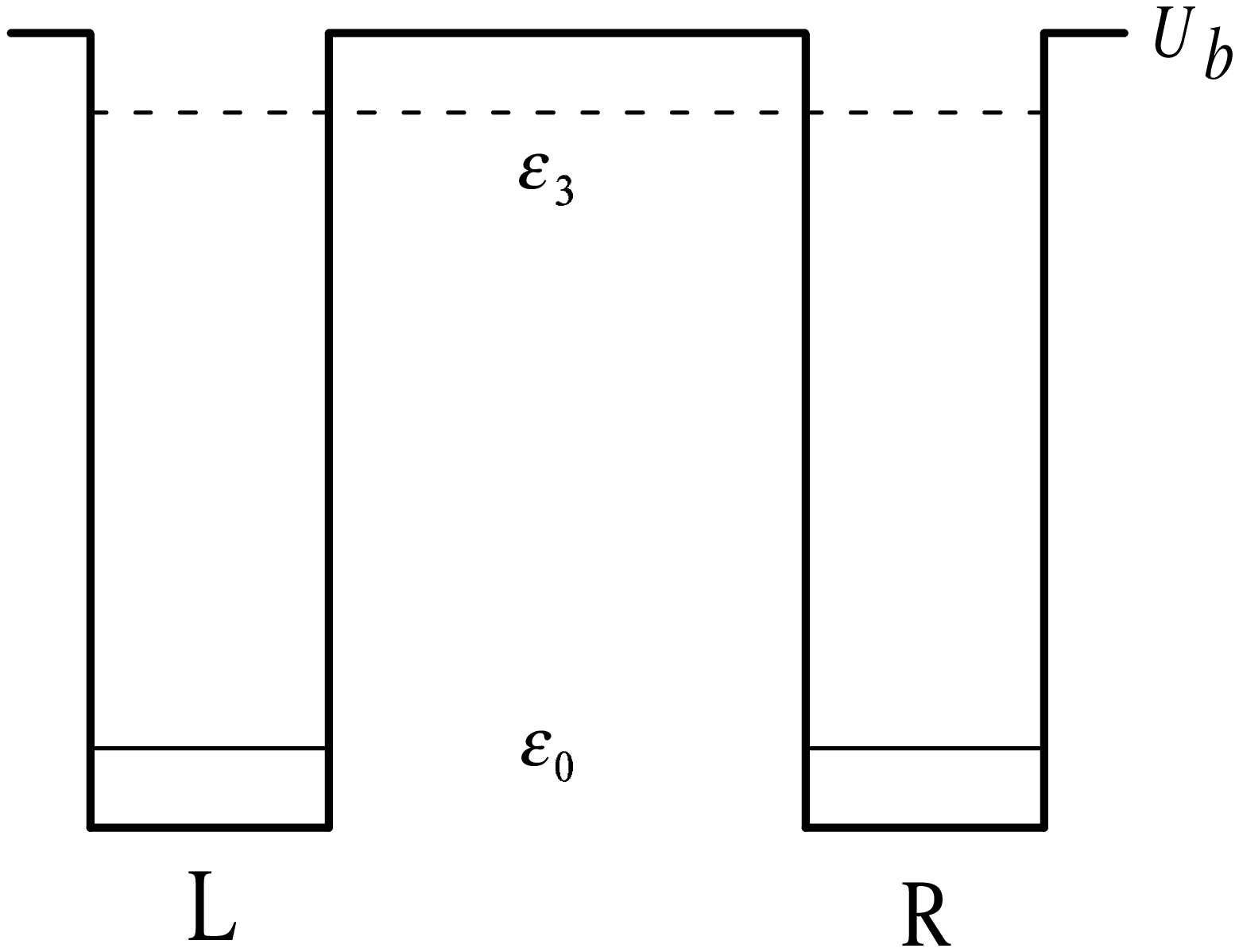}

\vskip 6mm

Fig. 1. Schematic energy diagram of the double-dot nanostructure.

\newpage

\includegraphics[width=\hsize]{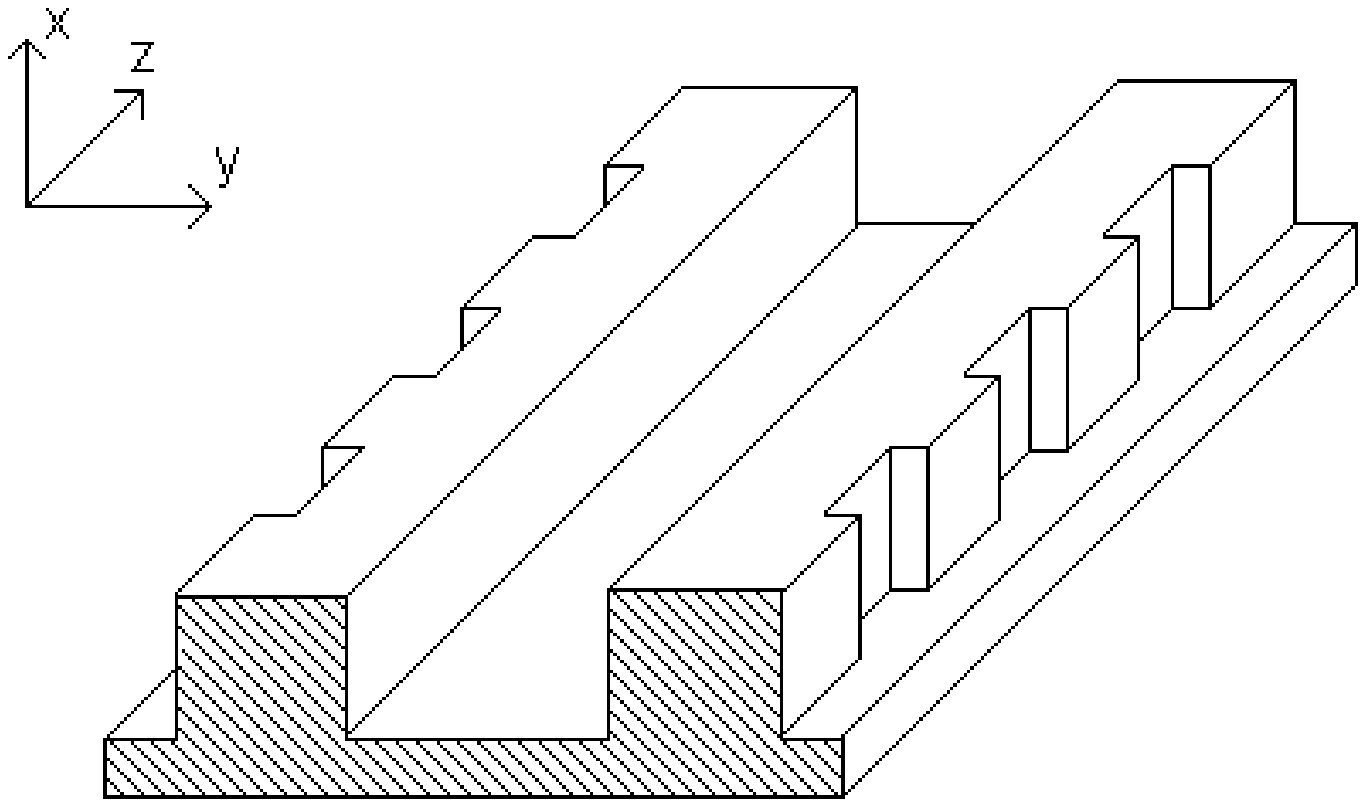}

\vskip 6mm

Fig. 2. Schematics of the double-waveguide structure studied in
Ref. \cite{Vorobeichik}.

\end{document}